# THE STELLAR MASS SPECTRUM IN THE YOUNG POPULOUS CLUSTER NGC 1866

Sidney van den Bergh
Dominion Astrophysical Observatory
Herzberg Institute of Astrophysics
National Research Council of Canada
5071 West Saanich Road
Victoria, British Columbia, V8X 4M6
Canada
vdb@dao.nrc.ca

## ABSTRACT

The young populous cluster NGC 1866 in the Large Magellanic Cloud LMC), which is probably one of the most massive object formed in the LMC during the last ~ 3 Gyr, appears to have an unexpectedly high mass-to-light ratio. From its velocity dispersion Fischer et al. (1992) find its mass to be $(1.35 \pm 0.25) \times 10^5$ $M_\odot$. The luminosity of this cluster is $M_V = -8.93 \pm 0.13$, corresponding to $L_V = (3.2 \pm 0.4) \times 10^5$ $L_V (\odot)$. This yields $M/L_V = 0.42 \pm 0.09$ in solar units. For a cluster of age 0.1 Gyr such a relatively high mass-to-light ratio requires a mass spectrum with an exponent $x = 1.72 \pm 0.09$; or $x = 1.75 \pm 0.09$ if mass loss by evolving stars is taken into account.

1.  **INTRODUCTION**

Just before the Second World War Baade (1963) asked Thackeray in South Africa to obtain photographs of the populous cluster NGC 1866 in the Large Magellanic Cloud. Surprisingly, Thackeray found that the brightest stars in this cluster were blue, not red, as might have been expected if NGC 1866 were a globular cluster. Subsequently Gascoigne & Kron (1952) showed that the brightest clusters in the LMC fall into two distinct color classes: (1) red globular clusters, and (2) populous blue clusters. In recent years (see Ashman & Zepf 1998 for numerous references) such populous blue clusters have been regarded as candidate young globular clusters. Van den Bergh & Lafontaine (1984) proposed that the existence of a number of populous blue star clusters in the Large Cloud was due to the fact that the mass spectrum of open clusters in the LMC is less deficient in high-mass clusters than is the case in the Galaxy and in M 31.

The LMC contains only five open clusters (van den Bergh & Hagen 1968) that are more luminous than NGC 1866. These are NGC 1818, NGC 1850, NGC 2004, NGC 2007 and NGC 2100. However, all of these clusters are, on the basis of their integrated colors, younger than NGC 1866. This cluster is therefore almost certainly the most massive young open cluster in the Large Cloud, and therefore one of the most massive cluster that formed in the LMC during the last ~ 3 Gyr.

By fitting of a CCD color-magnitude diagram of NGC 1866 to isochrones with Y = 0.28 and Z = 0.02, Brocato et al. (1989) find a cluster age of 110 Myr. A younger age of ~ 70 Myr is derived from the mean period, <log P(days)> = 0.49, of the eight Cluster Cepheids with HV numbers (Welch et al. 1991), and the period-age relation of Tammann (1969). In the subsequent discussion an age of 100 Myr will be adopted. The rate of star formation in the star field on which NGC 1866 is projected appears to exhibit a sharp increase ~ 3 Gyr ago (Stappers et al. 1997).

## 2. LUMINOSITY AND MASS OF NGC 1866

From photoelectric UBV photometry of NGC 1866 through a 60" aperture van den Bergh & Hagen (1968) obtained V(60) = 9.89 ± 0.01, B-V = 0.26 ± 0.02 and U-B = -0.06 ± 0.01. Furthermore these authors found V(42) = 10.26 through a 42" diaphragm. Using the outer part of the brightness profile of NGC 1866 observed by Fischer et al. (1992) the corresponding asymptotic magnitudes are V($\infty$) = 9.66 and V($\infty$) = 9.81, respectively. A value V($\infty$) = 9.74 ± 0.08 will be adopted. From UBV photometry the parameter Q = -0.25 (Johnson & Morgan 1953), yields $A_V$ = 0.17. With this value for the absorption $V_o$ = 9.57. In conjunction with an LMC distance modulus $(m-M)_o$ = 18.5 ± 0.1 (van den Bergh 2000) this yields $M_V$ = - 8.93 ± 0.13, corresponding to (3.2 ± 0.4) x $10^5$ $L_V$ ($\odot$) (Hayes 1985). Some uncertainty in the luminosity profile of NGC 1866, and hence in extrapolation to an asymptotic magnitude, results from the fact that the surface brightness which Fischer et al. (1992) obtain from star counts over the range 2 < R(pc) < 4 is an order of magnitude lower than that which these same authors derive from surface photometry; a difference which they attribute to the effects of crowding.

From the radial luminosity profile of, and the radial velocity dispersion in, NGC 1866 Fischer et al. (1992) derive the total cluster mass. Their analysis yields a total mass of $(1.35 \pm 0.25) \times 10^5$ $M_\odot$ for this cluster. From this mass, and the value $L_V = (3.2 \pm 0.4) \times 10^5$ $L_V (\odot)$ found above, one obtains a mass-to-light ratio $M/L_V = 0.42 \pm 0.09$. This value may be compared with the mass-to-light ratios computed by Bruzual & Charlot (1993), and Charlot (1999), for clusters of stars having solar composition and either a Scalo (1986) or a Salpeter (1955) mass spectrum with an upper mass cutoff at 126 $M_\odot$, and a lower mass cutoff at 0.1 $M_\odot$. Separate calculations were made for various slopes of the mass spectra, which were assumed to be of the form

$$dN/dm \propto m^{-(1+x)}, \qquad (1)$$

in which $x = 1.35$ is the value originally used by Salpeter. Interpolating between the values of $M/L_V$ given for T = 100 Myr in Table 1 one finds that $M/L_V = 0.42$

± 0.09 corresponds to x = 1.71 ± 0.09 assuming no mass loss, and x = 1.75 ± 0.09 when mass loss by evolving stars is taken into account. This is close to the upper limit for well-observed clusters for which x has been plotted by Scalo (1998). It would be interesting to know if the models by Charlot could be "tweaked" to reduce the difference between these values of x. Unfortunately the photometry by Brocato et al. (1989) does not extend deep enough to provide a direct determination of x for NGC 1866. Fisher et al. (1998) have, however, studied the luminosity function of the young LMC cluster NGC 2157, for which van den Bergh & Hagen (1968) measured V(60) = 10.16, B-V = 0.26, U-B = 0.06, from which Q = - 0.30. Their results show that NGC 2157 is slightly bluer (younger), and ~ 0.4 mag fainter, than NGC 1866. From Hubble Space Telescope (HST) Wide Field Planetary Camera observations down to V ~ 25.4 and I ~ 24.4 Fisher et al. (1998) find that x = $0.95^{+0.30}_{-0.25}$ in V and x = $1.0^{+0.25}_{-0.30}$ in I. In other words the luminosity function (and mass spectrum) of NGC 2157 are, if anything, slightly less steep than the standard Salpeter function. From HST observations of the even younger LMC cluster NGC 1818, Hunter et al. (1997) find x = 1.23 ± 0.08. These results suggest that the stars in NGC 1866 may have an unusually steep mass spectrum. Clearly it would be of great interest to obtain deep HST photometry of this cluster to confirm this conclusion. For a cluster with $M_V$ = - 8.93, an age of 100 Myr, and a Salpeter mass spectrum of slope x = 1.35 one would expect a total mass of 6.7 x $10^4$ $M_\odot$ without taking mass loss by evolving stars into account, or $5.6 \times 10^4$ $M_\odot$ with mass loss. These values are well below the mass of $(1.35 \pm 0.25) \times 10^5$ $M_\odot$ that Fischer et al. (1992) derived from the velocity dispersion and structural parameters of NGC 1866.

## 3. DISCUSSION

The $\sim 8 \times 10^4$ $M_\odot$ difference between the derived, and expected, masses of NGC 1866 might be explained in a number of different ways:

- The stellar mass spectrum of NGC 1866 might have a slope that is steeper than the canonical value $x = 1.35$. In this connection it is noted that Holtzman et al. (1997) have found that the luminosity function for faint field stars in the LMC may be steeper than that in the solar neighborhood. On the other hand, the well-observed young massive LMC clusters NGC 1818 and NGC 2157 have mass spectra that are shallower than $x = 1.35$.

- The missing mass might be in the form of a million or so brown dwarfs. Such objects would be very difficult to observe at the distance of the LMC.

- The missing mass could be in the form of tens of thousands of neutron stars. However, this would require an implausibly high supernova rate of a few per 10 000 years within NGC 1866.

- The true integrated luminosity of NGC 1866 is a few times greater than that which is derived from UBV photometry and its extrapolation to R($\infty$).

- The age of NGC 1866 is ~ 300 Myr, rather than 100 Myr, thus increasing the predicted value of $M/L_V$. However, such a large age cannot be reconciled with the cluster color-magnitude diagram, or with the presence of 3-day Cepheids.

- The dynamical mass determination is in error. The order of magnitude discrepancy between the surface brightness and star count data in the range 2 < R(pc) < 4 suggest that it might, perhaps, be desirable to revisit the dynamical mass determination for NGC 1866. It would clearly be very worthwhile to observe the luminosity distribution of NGC 1866 with the Hubble Space Telescope, since this would allow one to obtain a direct determination of the slope of its mass spectrum.

It is of interest to note that the mass of NGC 1866, which appears to be one of the most massive populous cluster ever formed in the Large Cloud, is only ~ 6 x $10^4$ $M_\odot$. This value is lower than that of ~ 90% of the Galactic globular clusters for which masses are known (Pryor & Meylan 1993).

It is a great pleasure to thank Stéphane Charlot for providing information on the evolution of clusters with steep mass spectra, and Philippe Fischer for helpful comments and suggestions.

# TABLE 1

# EVOLUTION OF MASSIVE STAR CLUSTERS[1,2]

| Spectrum | $M/L_V$ 10 Myr | $M/L_V$ 100 Myr | $M/L_V$ 14 Gyr |
|---|---|---|---|
| Scalo (1986) (24%) | 0.078 (2.2%) | 0.30 ( 6.7%) | 6 |
| x = 1.35[3] (32%) | 0.034 (7.4%) | 0.21 (17%) | 12 |
| x = 1.5 (23%) | 0.53 (3.9%) | 0.27 (11%) | 12 |
| x = 2.0 (6.6%) | 0.29 (0.0%) | 0.82 ( 1.2%) | 18 |
| x = 2.5 (1.3%) | 1.8 (0.0%) | 3.1 ( 0.0%) | 35 |
| x = 3.0 (0.0%) | 9.6 (0.0%) | 12.0 ( 0.0%) | 73 |

[1] Data kindly provided by Charlot (1999)

[2] Numbers in parentheses give amount by which $M/L_V$ has to be reduced to take mass loss into account

[3] Salpeter (1955)

- 11 -Salpeter, E.E. 1955, ApJ, 121, 161

Scalo, J.M. 1986, Fundam. Cosmic Phys., 11, 1

Scalo, J. 1998, in ASP Conference Series No. 142, The Stellar Initial Mass Function, Eds. G. Gilmore and D. Howell, (San Francisco: ASP), p. 201

Stappers, B.W. et al. 1997, PASP, 109, 292

Tammann, G.A. 1969, A&A, 3, 308

van den Bergh, S. 2000, The Galaxies of the Local Group, (Cambridge: Cambridge University Press), in press

van den Bergh, S. & Hagen, G.L. 1968, AJ, 73, 569

van den Bergh, S. & Lafontaine, A. 1984, AJ, 89, 1822

Welch, D.L., Mateo, M., Côté, P., Fischer, P. & Madore, B.F. 1991, AJ, 101, 490